\documentclass[showkeys,aps,prmat,reprint,superscriptaddress,onecolumn]{revtex4-1}
 \usepackage{epsfig}
 \usepackage{latexsym}
 \usepackage{epic}
 \usepackage{eepic}
 \usepackage{psfrag}
 \usepackage{rotating}
 \usepackage{multirow,booktabs}
 \usepackage{color}
 \usepackage{cprotect} 
 \usepackage{miller} 
 \usepackage{xspace} 

\usepackage{graphicx}
 
\usepackage{gensymb} 
 
  \usepackage[fleqn, leqno]{amsmath}
  \usepackage{amsfonts}  
  \usepackage{amssymb}   
 \usepackage{amsthm}
  \usepackage{stmaryrd} 
  \usepackage{import}

  \usepackage{array}

  \usepackage[normalem]{ulem}

\input{MyBibalias.tex}
\renewcommand\vec{\mathbf}

\newcommand{\ddr}{\ensuremath{{\mathrm d}{\mathbf r} }\xspace} 
\newcommand{\rr}{\ensuremath{{\mathbf r} }\xspace} 
\newcommand{\qq}{\ensuremath{{\mathbf q} }\xspace} 
 
\newcommand{\kk}{\ensuremath{{\mathbf k} }\xspace} 
\newcommand{\mm}{\ensuremath{{\mathbf m} }\xspace} 

\newcommand{\phii}[1][]{\ensuremath{\varphi}\xspace\ifx\relax#1\relax\else\ensuremath{\left(#1\right)}\xspace\fi}
 
\newcommand{\phiav}{\ensuremath{\bar{\phii}}\xspace} 
\newcommand{\BxO}{\ensuremath{B^x_0}\xspace} 
\newcommand{\DBO}{\ensuremath{\Delta B_0}\xspace}
\newcommand{\qo}{\ensuremath{q_0}\xspace} 

\newcommand{\FF}[1][]{\ensuremath{\mathcal{F}}\xspace\ifx\relax#1\relax\else\ensuremath{\left[#1\right]}\xspace\fi}
\newcommand{\FFm}[1][]{\ensuremath{\mathcal{F_{\mathrm m}}}\xspace\ifx\relax#1\relax\else\ensuremath{\left[#1\right]}\xspace\fi}
\newcommand{\FFpfc}[1][]{\ensuremath{\mathcal{F_{\mathrm{PFC}}}}\xspace\ifx\relax#1\relax\else\ensuremath{\left[#1\right]}\xspace\fi}
\newcommand{\FFcoup}[1][]{\ensuremath{\mathcal{F_{\mathrm c}}}\xspace\ifx\relax#1\relax\else\ensuremath{\left[#1\right]}\xspace\fi}

\newcommand{\ff}[1][]{\ensuremath{f}\ifx\relax#1\relax\else\ensuremath{\left(#1\right)}\fi}

\newcommand{\ffpfc}[1][]{\ensuremath{\ff_{\mathrm{PFC}}}\ifx\relax#1\relax\else\ensuremath{\left(#1\right)}\fi}
\newcommand{\ffid}[1][]{\ensuremath{\ff_{\mathrm{id}}}\ifx\relax#1\relax\else\ensuremath{\left(#1\right)}\fi}
\newcommand{\ffex}[1][]{\ensuremath{\ff_{\mathrm{ex}}}\ifx\relax#1\relax\else\ensuremath{\left(#1\right)}\fi}
\newcommand{\ffm}[1][]{\ensuremath{\ff_{\mathrm m}}\ifx\relax#1\relax\else\ensuremath{\left(#1\right)}\fi}
\newcommand{\ffcoup}[1][]{\ensuremath{\ff_{\mathrm c}}\ifx\relax#1\relax\else\ensuremath{\left(#1\right)}\fi}

\newcommand{\wwj}[1][]{\ensuremath{w(\kkj)}\ifx\relax#1\relax\else\ensuremath{\left({#1}\right)}\fi}
\newcommand{\NN}[1][]{\ensuremath{\mathcal{N}}\ifx\relax#1\relax\else\ensuremath{\left[{#1}\right]}\fi}
\newcommand{\NNj}[1][]{\ensuremath{\mathcal{N}_j}\ifx\relax#1\relax\else\ensuremath{\left[{#1}\right]}\fi}
\newcommand{\NNjp}[1][]{\ensuremath{\mathcal{N}_j^+}\ifx\relax#1\relax\else\ensuremath{\left[{#1}\right]}\fi}
\newcommand{\NNjm}[1][]{\ensuremath{\mathcal{N}_j^-}\ifx\relax#1\relax\else\ensuremath{\left[{#1}\right]}\fi}

\newcommand{\LL}[1][]{\ensuremath{\mathcal{L}}\ifx\relax#1\relax\else\ensuremath{#1}\fi}
\newcommand{\LLj}[1][]{\ensuremath{\mathcal{L}_j}\ifx\relax#1\relax\else\ensuremath{#1}\fi}
\newcommand{\GGj}[1][]{\ensuremath{\mathcal{G}_j}\ifx\relax#1\relax\else\ensuremath{#1}\fi}
\newcommand{\QQj}[1][]{\ensuremath{\mathcal{Q}_j}\ifx\relax#1\relax\else\ensuremath{#1}\fi}

\newcommand{\np}[1][]{\ensuremath{#1}\ifx\relax#1\relax\else\ensuremath{^{n+1}}\fi}
\newcommand{\n}[1][]{\ensuremath{#1}\ifx\relax#1\relax\else\ensuremath{^{n}}\fi}

\newcommand \be {\begin{eqnarray}}
\newcommand \ee {\end{eqnarray}}

\newcommand{\kkj}{\ensuremath{\vec{k}_j}\xspace}
\newcommand{\DD}{\ensuremath{\vec{D}}\xspace}
\newcommand{\DDk}{\ensuremath{\vec{D}_{\mathrm k}}\xspace}

\newcommand{\sof}[1][]{\ensuremath{\{#1\}}\xspace}
\newcommand{\sofk}{\ensuremath{\sof[\kkj]}\xspace}

\newcommand{\Aj}{\ensuremath{A_j}\xspace}
\newcommand{\A}{\ensuremath{A}\xspace}

\newcommand{\Ajcc}{\ensuremath{A^*_j}\xspace}

\newcommand{\sofA}{\ensuremath{\sof[\Aj]}\xspace}
\newcommand{\sofAcc}{\ensuremath{\sof[\Ajcc]}\xspace}

\newcommand{\meS}[1][]{\ensuremath{\mathrm S}\ifx\relax#1\relax\else\ensuremath{_{#1}}\fi}

\newcommand{\MES}{MES\xspace}
\newcommand{\MM}[1][]{\ensuremath{\mathcal{M}}\ifx\relax#1\relax\else\ensuremath{#1}\fi}
\newcommand{\MC}[1][]{\ensuremath{\mathcal{C}}\ifx\relax#1\relax\else\ensuremath{#1}\fi}
\newcommand{\MMj}[1][]{\ensuremath{\mathcal{M}_j}\ifx\relax#1\relax\else\ensuremath{#1}\fi}
\newcommand{\MMaj}[1][]{\ensuremath{\mathcal{M}_{1,j}}\ifx\relax#1\relax\else\ensuremath{#1}\fi}
\newcommand{\MMbj}[1][]{\ensuremath{\mathcal{M}_{2,j}}\ifx\relax#1\relax\else\ensuremath{#1}\fi}
\newcommand{\MCj}[1][]{\ensuremath{\mathcal{C}_j}\ifx\relax#1\relax\else\ensuremath{#1}\fi}
\newcommand{\ii}{\ensuremath{\mathrm{\mathbf{i}}}\xspace}

\newcommand{\aalpha}{\ensuremath{\alpha_2}\xspace}

\newcommand*\colvec[3][]{
    \begin{pmatrix}\ifx\relax#1\relax\else#1\\\fi#2\\#3\end{pmatrix}
  }
\newcommand{\mycvec}[1]{\ensuremath{\begin{pmatrix}#1\end{pmatrix}}}
  
\begin{document}

\title{Controlling magnetic anisotropy in amplitude expansion of phase field crystal model}


\author{Rainer Backofen}\email[]{rainer.backofen@tu-dresden.de}
\affiliation{
  Institute of Scientific Computing, Technische Universit\"at Dresden, 01062 Dresden, Germany}
\author{Marco Salvalaglio}
\affiliation{
  Institute of Scientific Computing, Technische Universit\"at Dresden, 01062 Dresden, Germany}
\affiliation{Dresden Centre for Computational Materials Science (DCMS), TU Dresden, 01062 Dresden, Germany}
\author{Axel Voigt}
\affiliation{
  Institute of Scientific Computing, Technische Universit\"at Dresden, 01062 Dresden, Germany}
\affiliation{Dresden Centre for Computational Materials Science (DCMS), TU Dresden, 01062 Dresden, Germany}

\begin{abstract}
The amplitude expansion for a magnetic phase-field-crystal (magnetic APFC) model enables a convenient coarse-grained description of crystalline structures under the influence of magnetic fields. Considering higher-order magnetic coupling terms, we demonstrate the possibility of tuning the magnetic anisotropy in these models. This allows for reproducing the easy and hard direction of magnetization. Such a result can be achieved without increasing the computational cost, enabling simulations of the manipulation of dislocation networks and microstructures in ferromagnetic materials. As a demonstration, we report on the simulation of the shrinkage of a spherical grain with the magnetic anisotropy of Fe. 
\end{abstract}

\maketitle

\section{Introduction}

In order to explore the possibilities external magnetic fields offer to manipulate microstructure in ferromagnetic materials \cite{Guillonetal_MT_2018}, a detailed understanding of the interactions between magnetic fields and solid-state matter transport is required. Various modeling approaches have been introduced to describe the magneto-structural interactions in a multiscale framework. A promising approach is the phase-field-crystal (PFC) model \cite{EKG02,EG04}, describing crystal lattices through a periodic density field, that was extended to capture the fundamental physics of magnetocrystalline interactions \cite{FPK13,SSP15}. In \cite{FPK13}, the PFC density is coupled with magnetization to generate a ferromagnetic solid below a critical temperature, while in \cite{SSP15} this PFC model is extended to multiferroic binary solid solutions and used to demonstrate the influence of magnetic fields on the growth of crystalline grains. Magneto-structural interactions are incorporated phenomenologically and building on symmetry arguments. This model, which consists of a system of evolution equations for the rescaled atomic density field $\phii$ and an averaged magnetization $\mathbf{m}$, is used in \cite{BEV19,BV20} in a simplified form to study the role played by external magnetic fields on the evolution of defect structures, grain boundaries, long-time scaling behaviors and various geometrical and topological properties in grain growth. While the microscopic details are well resolved with the considered magnetic PFC model and experimental relevant time scales can be reached, the required spatial resolution restricts simulations to two-dimensional settings. Furthermore, all previous investigations have considered generic material parameters for magnetic anisotropy. 

The complex amplitude PFC (APFC) model initially introduced in \cite{GAD05,AGD06} provides a framework to overcome the restriction resulting from the spatial resolution required by PFC models. The idea is to model the amplitude of the density fluctuations instead of the density itself. This allows for reaching larger spatial scales while retaining essential microscopic effects \cite{Spatschek2010,EHP10,SVE19}, thus enabling mesoscale investigations of crystalline systems. For a recent review of APFC models, we refer to \cite{Salvalaglio2022}. In \cite{Backofen2022}, a magnetic APFC model is introduced, and the applicability for a simple three-dimensional setting has been demonstrated. Together with advanced numerical approaches \cite{PSV19},
this enables the description of magneto-structural interactions in multiscale simulations, combining the dynamics of defects, dislocation networks, and grain boundaries with experimentally accessible microstructure evolution on diffusive time scales \cite{Salvalaglio2022}.
We here build on this approach and modify the magnetic coupling energy. The considered modification allows for tuning the magnetic anisotropy, reproducing the easy and hard direction of ferromagnetic materials. We demonstrate this for BCC and FCC crystals. This modification essentially overcomes the limitations of previous approaches and enables the modeling of material-specific magnetic anisotropies. 

The paper is structured as follows: In Section \ref{sec:2}, we briefly review the magnetic PFC and APFC models. We describe the
numerical approach to solve the magnetic APFC model, discuss a modification of the magnetic coupling energy which does not increase the computational cost, and introduce the few-mode approximation and the minimal energy surface in the reciprocal space as a tool to analyze the impact of the magnetic coupling. In Section \ref{sec:3}, we analyze the magnetic properties of BCC and FCC crystals and demonstrate the possibility of tuning the easy and hard direction with the modified magnetic coupling energy. We further apply this new setting to study the magnetic impact on grain growth. In Section \ref{sec:4} we draw conclusions.

\section{Magnetic PFC and APFC models} \label{sec:2}

\subsection{Magnetic PFC model}
In \cite{FPK13,SSP15} a magnetic PFC model was proposed. This model describes the basic phenomenology of magneto-structural interactions in crystals, namely magnetic anisotropy, and magneto-striction. In the limit of constant magnetization or strong external magnetic field, the free energy, on which the magnetic PFC model builds, reads
\begin{align}
  \FF[\phii,\mm]&=\int_{\Omega} \bigg[ \frac{B^x_0}{2} \phii(\qo^2+\nabla^2)^2\phii
  +\frac{\Delta B_0}{2}\phii^2
  -\frac{t}{3}\phii^3+\frac{v}{4}\phii^4 \bigg] \ddr + \FF_\mm[\phii, \mm], 
   \label{eq::cntrPFCenergy} \\
  \FF_\mm[\phii,\mm]&=\int_{\Omega} \bigg[ \widetilde{\alpha}_0 \mm^2 \phii^2 + \sum_{i=1} \frac{\widetilde{\alpha}_{2i}}{2i} (\mm \cdot \nabla \phii)^{2i} \bigg] \ddr,
  \label{eq::cntrPFCenergyMag}
\end{align}
where $\phii$ denotes the scaled particle density and $\mathbf{m}$ the magnetization. $\qo$ defines the lattice spacing at equilibrium, $\Omega$ is the domain of integration, $\BxO$, $\DBO$, $\tau$ and $v$ are parameters as introduced in \cite{Elder2007}. Together with the average density $\phiav$, they define crystal structure and physical properties. $\FF_\mm[\phii,\mm]$ accounts for the magnetocrystalline interactions with magnetization \mm, which is assumed to be constant and scaled to unit length, $|\mathbf{m}| \,= 1$. Even powers are considered in the expansion due to the required mirror symmetry $\mm \to - \mm$. The parameters $\widetilde{\alpha}_{2i}$ can be tuned to control the magnetic anisotropy. However, already by setting $\widetilde{\alpha}_0 = 0$ and considering the expansion only to lowest order, resulting in $\FF_\mm[\phii,\mm] = \int_\Omega \frac{\widetilde{\alpha}_2}{2} (\mm \cdot \nabla \phii)^2 \ddr$, leads to magnetic anisotropy and can be considered as a minimal model. In this setting, $\widetilde{\alpha}_2$ simply controls the strength of the magnetic interaction. Integrating by parts allows to rewrite this as $\FF_\mm[\phii,\mm] = \int_\Omega - \frac{\widetilde{\alpha}_2}{2} \phii (\mm \cdot \nabla)^2 \phii \ddr$, which is numerically advantageous and has been considered in \cite{BEV19,BV20}. The evolution equation for $\varphi$ reads
\begin{equation}
    \frac{\partial \varphi}{\partial t}=\nabla^2 \frac{\delta \mathcal{F}}{\delta \varphi}.
    \label{eq::PFCdynamics}
\end{equation}
As the expansion in eq.~\eqref{eq::cntrPFCenergyMag} is phenomenological, we can also propose a different expansion fulfilling the same symmetry constraints
\begin{equation}
    \FF_{\mm^\prime}[\phii,\mm]=\int_{\Omega} \bigg[ \hat{\alpha}_0 \mm^2 \phii^2 + \sum_{i=1} \frac{\hat{\alpha}_{2i}}{2} \phii (\mm \cdot \nabla)^{2i} \phii \bigg] \ddr.
  \label{eq::cntrPFCenergyMagPrime}
\end{equation}
We again set $\hat{\alpha}_0 = 0$ and, as already seen, in lowest order $\FF_\mm[\phii,\mm] = \FF_{\mm^\prime}[\phii,\mm]$ if $\hat{\alpha}_2 = - \widetilde{\alpha}_2$. However, in this formulation also higher-order terms can be considered in a numerically efficient manner. We will consider expansions to order $i = 2$ and demonstrate the possibility of tuning the magnetic anisotropy.

\subsection{Magnetic APFC model}
In the crystal phase, the density field \phii described by the PFC model is periodic, with maxima at the atomic-lattice sites, thus encoding the crystal structure directly. In the amplitude expansion of the PFC model, this density is expanded in terms of a small set of Fourier modes
\begin{equation}
 \phii(\rr) = \phiav+\sum_{j=1}^{N}\left[ \Aj(\rr) {\rm e}^{\ii \kkj \cdot \rr} +  \Aj^*(\rr) {\rm e}^{-\ii \kkj \cdot \rr} \right],
\label{eq::APFCref}
\end{equation}
with \phiav the overall mean density and $\sofk_{j=1}^N$ defining the symmetry of a reference crystal usually corresponding to a bulk, relaxed lattice. The reference crystal is then described by real and constant amplitudes $\Aj \in \mathbb{R}$. Complex and space-dependent amplitudes account for deviations from the reference crystal. $|\Aj|$ entails information on the local ordering while the phase of the complex amplitudes accounts for displacement with respect to the reference crystal.
Thus, at defects where singular displacement occurs, some of the amplitudes vanish, namely the ones having singular phases. Far from defects, both phase and argument of \Aj vary typically on a larger length scale than the distance between particles in the crystal. Only at defects the amplitudes vary at a similar scale. 

In \cite{Backofen2022} a corresponding magnetic APFC model to eqs. \eqref{eq::cntrPFCenergy}, \eqref{eq::PFCdynamics} and \eqref{eq::cntrPFCenergyMagPrime} has been derived. It results from substituting eq.~\eqref{eq::APFCref} into eq.~\eqref{eq::cntrPFCenergy} and \eqref{eq::cntrPFCenergyMagPrime} and averaging fluctuations on small scales. The resulting equations read
\begin{equation}
  \FF[\sofA]=\int_{\Omega} \bigg[
                    \sum_{j=1}^N  B_0^x \left( \Ajcc \GGj^2 \Aj
                      + \Ajcc \MCj \Aj \right)+ g^{\rm S}(\sofA)\bigg] \ddr,
                    \label{eq::APFCenergy}
\end{equation}
with
\begin{equation}
g^{\rm S}(\sofA) = \sum_{j=1}^N  \left( -\frac{3v}{2}|\Aj|^4 \right)
                 + \frac{\Delta B_0}{2}A^2
                 + \frac{3v}{4}A^4
                 +f^{\rm S}(\sofA), \label{eq:APFCgS}
               \end{equation}
 and $A^2\equiv 2\sum_{j=1}^N |\Aj|^2$, $A^4 \equiv (A^2)^2$, $\GGj \equiv
 q_0^2-|\kkj|^2+\nabla^2+2\ii\,\kkj \cdot \nabla$ and $\MCj = \frac{1}{\BxO} \sum_{i=1} \hat{\alpha}_{2i} (\mm \cdot \nabla + \ii \mm \cdot \kkj)^{2i}$. 
 
 The magnetic coupling is considered in the terms $\Ajcc \MCj \Aj$. In \cite{Backofen2022} only the lowest order ($i = 1$) is considered, leading to a minimal magnetic APFC model. The equilibrium crystal without magnetization is chosen as a reference, which leads to $|\kkj| = \qo$. $f^{\rm S}$ is a polynomial function in \sofA and \sofAcc. It depends on the reference crystal structure, see \cite{EHP10,SBE17,Salvalaglio2022}.
 
 The evolution equations for each amplitude read
 \begin{equation} \label{eq:APFC}
  \frac{\partial \Aj}{\partial t}
  = -|\mathbf{k}_j|^2 \frac{\delta \FF}{\delta \Ajcc},
  \end{equation}
with
\begin{equation}
  \frac{\delta \FF}{\delta \Ajcc} = B_0^x\left[ \mathcal{G}_j^2 + \MCj \right]\Aj + \frac{\partial g^{\rm S}(\sofA)}{\partial \Ajcc}.
\label{eq::APFCevol}
\end{equation}

Considering $\MCj = \MMj^2 + \QQj$ we can write 
\begin{align}
  \left[\GGj^2+ \MM_j^2 \right]
  =(\GGj+ \ii \MMj)(\GGj- \ii \MMj)
  =:\NNjp \NNjm,
\end{align}
and eqs. \eqref{eq:APFC} can be written as systems of second-order equations
\begin{equation}
\begin{split} \label{eq:mAPFC}
  \frac{\partial \Aj}{\partial t} &=  -|\mathbf{k}_j|^2 \left[ B_0^x \NNjp \mu_j + B_0^x \QQj \Aj + G_j(\sofA) \right], \\
  \mu_j&= \NNjm \Aj,
\end{split}
\end{equation} 
with $G_j(\sofA) := {\partial g^{\rm S}(\sofA)}/{\partial \Ajcc}$ the nonlinear terms. In the following, we consider two specific forms of the magnetic coupling term:
\begin{align}
 \label{eq:MA}  \text{Model A} : \quad \MMj^2&= \aalpha \left(\mm \cdot \nabla+ \ii\,\mm \cdot \kkj \right)^2  \quad \text{and} \quad \QQj=0, \\
 \label{eq:MB} \text{Model B} : \quad \MMj^2&= \aalpha \left[q_m^2+(\mm \cdot \nabla + \ii\,\mm \cdot \kkj)^2\right]^2 \quad \text{and} \quad \QQj=-\aalpha\, q_m^4,
\end{align}
with $\aalpha = \hat{\alpha}_2/B_0^x$. Model A corresponds to the lowest order expansion $i = 1$ and has been considered in \cite{Backofen2022} and Model B considers also the next higher order term $i = 2$. It is only reformulated by completing the square. It introduces an additional parameter $q_m$ which can be used to tune the magnetic anisotropy. The original parameters are obtained by $\hat{\alpha}_2 = 2 \alpha_2 q_m^2 B_0^x$ and $\hat{\alpha}_4 = \alpha_2 B_0^x$. {Model A} and {Model B} can be solved with almost the same computational cost. We follow the numerical approach in \cite{Backofen2022}, see \cite{SBE17,PSV19} for further details. The FEM discretization is implemented in the parallel and adaptive finite element toolbox AMDiS \cite{VV07,Witkowski2015}. 

\subsection{Minimum Energy Surface (\MES)}
\label{sec::cntrFMA}
In order to analyze the influence of $\mm$ in a bulk system, it is not necessary to solve eqs. \eqref{eq:mAPFC}. For a single crystal without defects and deformations, which are constant in space, the density can be expanded as in eq.~\eqref{eq::APFCref} considering deformed reciprocal space vectors $\kkj^\prime = \DDk \kkj$, with $\DDk=(\DD^{-1})^{\rm T}$ and $\DD$ the deformation matrix, see \cite{Backofen2022}. This description is exact for homogeneous deformed single crystals and provides a good approximation in more general situations \cite{JA10b}. For Model A, the free energy, eq.~\eqref{eq::APFCenergy}, simplifies to
\begin{align}
  \FF[\sofA,\{\kkj^\prime\}]= |\Omega^\prime| \Big[\sum_{j=1}^N \Aj \underbrace{B_0^x\left[(q_0^2-{\kkj^\prime}^2)^2 - 
  \alpha_{2} (\mm \cdot \kkj^\prime)^2 \right]}_{w(\kkj^\prime)} \Aj + g^{\mathrm{S}}(\sofA) \Big], 
  \label{eq::FMAenergy}
\end{align}
where $|\Omega^\prime|$ is the volume of the integration domain. The elastic and magnetic properties of the model are solely governed by the kernel $w(\mathbf{q})$. The first term in $w(\mathbf{q})$ corresponds to the approximation of the excess free energy of classical density functional theory \cite{EPG07,TBL09,ARRS19} and resembles the approximation of a correlation function in the reciprocal space. It is invariant on the orientation of $\mathbf{q}$ and, thus, on the orientation of the crystal structure. The rotational symmetry of the crystal is broken by the magnetization, \mm. This is reflected in the second term of $w(\mathbf{q})$, which depends on the relative orientation of \mm and
$\mathbf{q}$ and the coupling strength $\alpha_2$. 

Model B leads to a similar expression with
\begin{align}
w(\kkj^\prime) = B_0^x\left[(q_0^2-{\kkj^\prime}^2)^2 + 
  \aalpha (q_m^2-(\mm \cdot \kkj^\prime)^2)^2\right].
  \label{eq::FMAenergyKernalB}
\end{align}

\begin{figure}[htb]
  \noindent
 \begin{tabular}{c|ccc}
        \multicolumn{1}{c|}{no magnetic coupling} & \multicolumn{3}{c}{magnetic coupling} \\
    \multicolumn{1}{c|}{isotropic} & \multicolumn{2}{c}{anisotropic} & \\
    \multicolumn{1}{c|}{} &Model A & Model B & \\
    \includegraphics*[width =0.2 \textwidth]{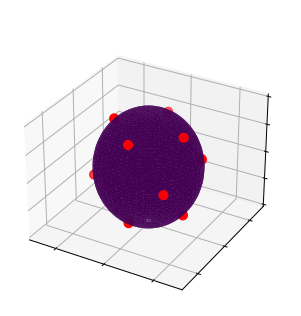} &
    \includegraphics*[width = 0.2 \textwidth]{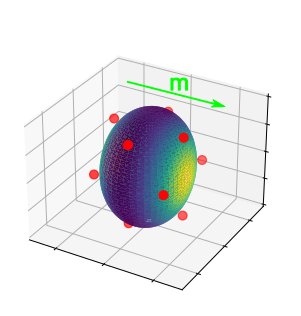} &
    \includegraphics*[width = 0.2 \textwidth]{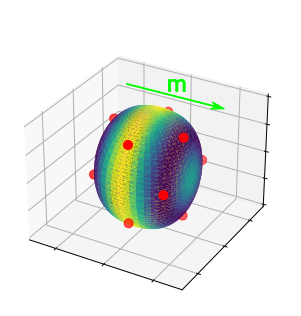} & 
                                                                          \includegraphics*[width = 0.02 \textwidth]{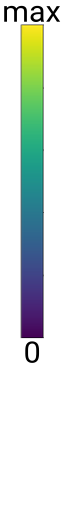} \\
   \multicolumn{1}{l}{a)} &\multicolumn{1}{l}{b)} &\multicolumn{1}{l}{c)} & \\
  \end{tabular}  
  \begin{center}
\begin{minipage}{0.95\textwidth}
\caption[short figure description]{
Symmetry breaking due to magnetic interaction. Minimum energy surface (\MES) without and with magnetic coupling. The red spheres corresponds to the (shortest) \kkj's representing an undeformed BCC crystal. The color indicates the energy contribution in the reciprocal space of \wwj. For $\aalpha > 0$, the energy is increased in the direction of \mm $\parallel \hkl[100]$, and the \MES results comporessed in this direction. In Model A the MES becomes ellipsoidal and the \kkj's are no longer on the deformed MES. In Model B, $q_m=1/\sqrt{2}$ is chosen and all the \kkj's are still on the MES.  
Here the deformation of \MES is exaggerated by considering unrealistic large $\aalpha = -0.75$ (Model A) and  $\aalpha = 0.75$ (Model B) for illustration purposes. 
\label{fig::MES}
}
\end{minipage}
\end{center}
\end{figure}

To visualize the impact of the magnetization on the crystal structure
we consider the energy contribution given by $w(\mathbf{q})$. For a given orientation $\mathbf{q}$ in the reciprocal space, $w(\mathbf{q})$ is minimized by adapting the length of $\mathbf{q}$. The resulting lengths for all possible $\mathbf{q}$'s can be plotted as a surface in reciprocal space, referred as Minimum Energy Surface (\MES). Fig.~\ref{fig::MES} shows such surfaces for different cases with a color map corresponding to the energy contribution from $w(\mathbf{q})$. In Fig.~\ref{fig::MES}~a) the contribution of the first term in $w(\mathbf{q})$ is shown, corresponding to the case without magnetization. Considering only this part, $\aalpha = 0$, a monochrome sphere with radius $q_0$ is obtained. The $\kkj^\prime$ vectors lie on this sphere, as shown for a BCC crystal by the red dots in Fig.~\ref{fig::MES}~a) (see Fig.~\ref{fig::kjs} for details), and any rotation of the crystal does not change the energy. This reflects the rotational symmetry of the model inherited from the correlation function. Deformations shift the vectors $\kkj^\prime$ away from the \MES, thus leading to an increase in the energy. 
When considering the magnetic coupling, the scenario changes. 
The energy depends on the orientation with respect to \mm. For directions aligned with \mm, $w(\mathbf{q})$ is increased for $\aalpha < 0$. In addition, the \MES is deformed. For $\aalpha < 0$ it shrinks in the direction of \mm. The \kkj vectors describing a relaxed crystal then do not lie anymore on the \MES. This leads to additional effects such as the tendency of the crystal to deform in order to minimize the energy, or in other words the deviation from the MES, known as magnetostriction. We note that this effect is relatively small. As the $\kkj^\prime$ vectors cannot vary independently they cannot always lie on the \MES, as seen in Fig.~\ref{fig::MES}~b), where the effect is exaggerated with large values of $\alpha_2$ for illustration purposes. As discussed below, however, parameters entering Model B can be tuned to have both $\kkj$ and $\kkj^\prime$ on the \MES (thus suppressing magnetostriction). With all the information it conveys, the \MES can then be used as a suitable tool to study the impact of magnetic coupling.


\section{Analysis and Simulation} \label{sec:3}
\subsection{Magnetic Properties}
We consider eq.~\eqref{eq::FMAenergy} to calculate the free energy of a single crystal. We minimize the free energy w.r.t
\sofA and compute the deformation 
along and perpendicular to \mm, this leads to the estimation of the eigenvalues of the deformation matrix, $d_0$ and $d_1$, see \cite{Backofen2022}. This deformation defines the magnetostriction. The energy dependence on the direction of \mm defines the magnetic
anisotropy. The direction with the lowest and highest energy are called easy and hard direction of magnetization, respectively. BCC as well as FCC crystals are considered. 

\begin{figure}[htb]
  \noindent
  \begin{center}
    \includegraphics*[angle = -0, width = 0.45 \textwidth ]{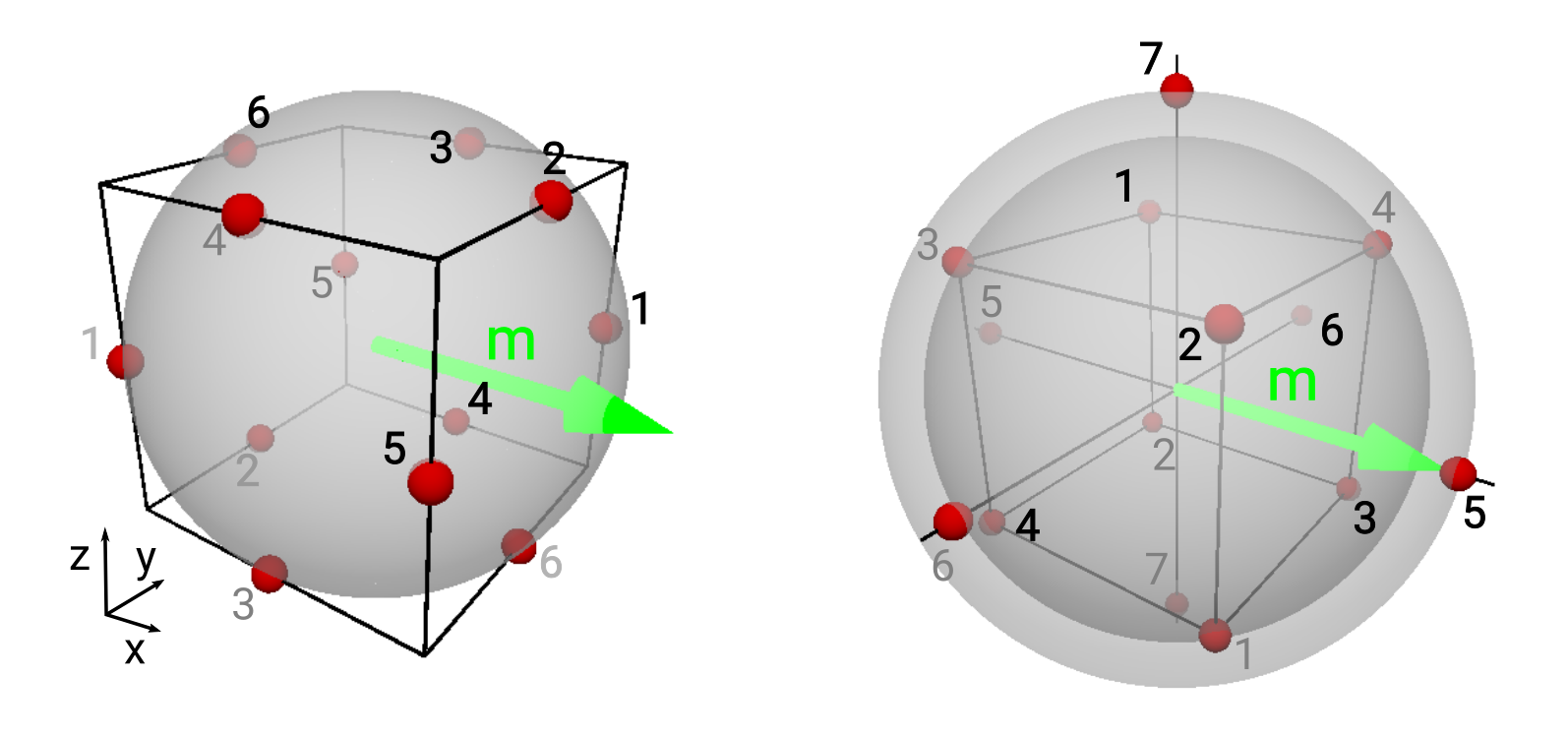}
\begin{minipage}{0.95\textwidth}
\caption[short figure description]{
  Crystal structure. The red spheres represent the $\kkj$ vectors. Their numbering is according to eqs. \eqref{eq::kjsBCC} and \eqref{eq::kjsFCC}. The energetic equivalent
  vectors, $-$\kkj, are numbered in grey. BCC (left) and FCC (right) crystals are illustrated. \mm is shown in green and is aligned to the \hkl[100] direction.
\label{fig::kjs}
}
\end{minipage}
\end{center}
\end{figure}

\subsubsection{BCC crystals}
For BCC crystals the structure dependent part in eq.~\eqref{eq:APFCgS} reads
\begin{align}
 f^{\mathrm{BCC}}(\sofA)=& -2t(\A_1^*\A_2\A_4+\A_2^*\A_3\A_5+\A_3^*\A_1\A_6
                               + \A_4^*\A_5^*\A_6^*+{\rm c.c.})  \nonumber \\
                         &+6v(\A_1\A_3^*\A_4^*\A_5^*+\A_2\A_1^*\A_5^*\A_6^*
  +\A_3\A_2^*\A_6^*\A_4^*+{\rm c.c.})  \label{eq::gBCC}
\end{align}
with $\kkj$ vectors defined as, see e.g. \cite{SBE17},
\begin{align}
  \sof[\kkj]&= \frac{1}{\sqrt{2}} \left[ \mycvec{1 \\ 1 \\ 0},   \mycvec{1\\0\\1},  \mycvec{0\\1\\1}, \mycvec{0\\1\\-1},  \mycvec{1\\-1\\0},  \mycvec{-1\\0\\1} \right] 
  \label{eq::kjsBCC} 
\end{align}
where we have used $q_0 = 1$. They are numbered from $j = 1, \ldots, 6$ and shown in Fig.~\ref{fig::kjs}(left).

\begin{figure*}[htb]
  \noindent
  \centering
  \begin{tabular}{cc}
     \multicolumn{1}{l}{a)} & \multicolumn{1}{l}{b)}  \\
    \includegraphics*[width = 0.32 \textwidth] {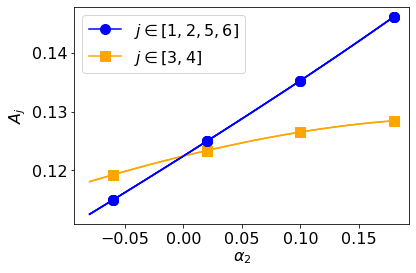} 
    & \includegraphics*[width = 0.32 \textwidth] {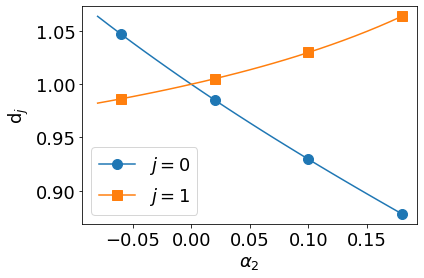} \\
    \multicolumn{1}{l}{c)} & \multicolumn{1}{l}{d)}  \\
    \includegraphics*[width = 0.32 \textwidth] {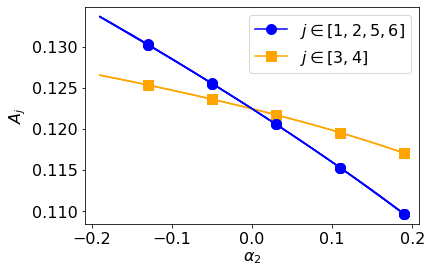} 
    & \includegraphics*[width = 0.32 \textwidth] {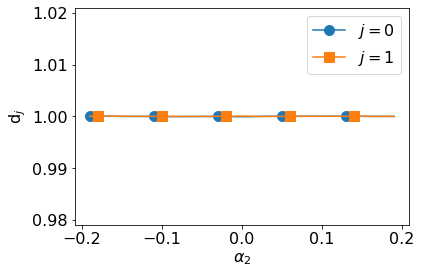} \\
   \end{tabular}
   \begin{center}
\begin{minipage}{0.9\textwidth}
  \caption[short figure description]{\label{fig:MESBCC} Comparison of coupling methods for a BCC crystal and magnetization along the \hkl[100] direction in terms of $\{A_j\}$ and $d_j$ as function of $\alpha_2$: a)-b) Model A and c)-d) Model B (with $q_m=1/\sqrt{2}$).
  }
\end{minipage}
\end{center}
\end{figure*}

In Fig.~\ref{fig:MESBCC} we compare Model A and Model B for \mm in \hkl[100] direction and different \aalpha. Fig.~\ref{fig:MESBCC}~a) and b) consider Model A and show in panel a) the minimized amplitudes \sofA as a function of \aalpha and in panel b) the deformations along and perpendicular to \mm, $d_0$ and $d_1$, for different \aalpha. Fig.~\ref{fig:MESBCC}~c) and d) show the same quantities for Model B with $q_m = 1/\sqrt{2}$. For \aalpha = 0, all amplitudes are equal, and there is no deformation. However, with magnetic coupling, differences occur. The amplitudes \Aj depend on \aalpha according to the relative orientation of the corresponding $\kkj$ vectors to \mm. For those which are perpendicular, the influence is less pronounced. In these directions, the \MES is not influenced by magnetic coupling. All others contribute equally. This behavior is qualitatively the same for Model A and Model B, but with opposite signs. Differences between the models are found in the deformation. While Model A leads to an expansion in the \mm direction for negative \aalpha and the opposite behavior perpendicular to \mm, there is no deformation in Model B. This difference results from the double well structure of $w(\qq)$ of Model B and the choice of $q_m=1/\sqrt{2}$. The latter is indeed chosen to have $\kkj$ lying on the \MES and thus to have no deformation ($d_0=d_1=1$) for $\kkj'$, independently of \aalpha. Model B then allows for decoupling magnetic anisotropy from magnetostriction effects and, in general, tuning these properties through $q_m$ and $\alpha_2$.

\begin{figure*}[htb]
  \noindent
  \centering
\begin{tabular}{cccc}
  \multicolumn{4}{l}{a)} \\
    \multicolumn{4}{c}{\includegraphics*[width = 0.35 \textwidth]{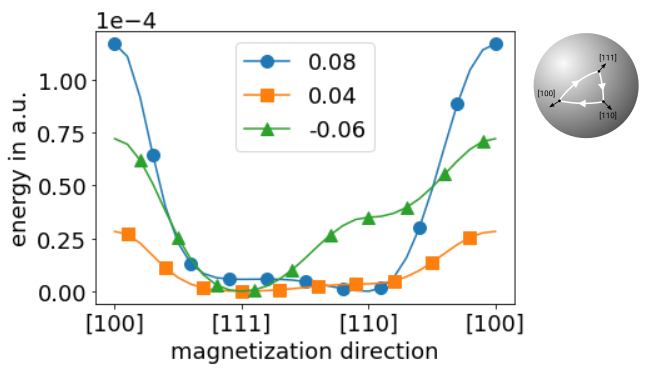}} \\
  \multicolumn{1}{l}{b)} & \multicolumn{1}{l}{c)} & \multicolumn{1}{l}{d)} & \\

\includegraphics*[width=0.22\textwidth] {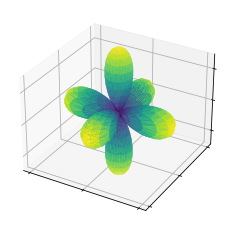} &
\includegraphics*[width=0.22\textwidth] {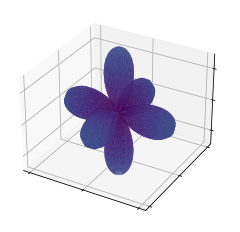} &
\includegraphics*[width=0.22\textwidth]{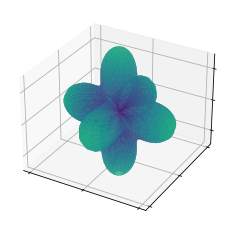}  
& \includegraphics*[width=0.02\textwidth]{nice_plots/colorbar}\\
\aalpha=0.08 & \aalpha=0.04 & \aalpha=-0.06 &\\
\end{tabular}
\begin{center}
\begin{minipage}{0.9\textwidth}
\caption[short figure description]{
BCC, Model A. Comparing magnetic anisotropy for \aalpha = 0.08, 0.04 and -0.06. a) Energy along the edges of the triangle defined by \hkl[100], \hkl[111] and \hkl[110]. Maximum defines the hard direction and minimum the easy direction of magnetization. The energy is plotted relative to the minimum energy (easy direction). b)-d) Energy surface for \aalpha = 0.08, 0.04 and -0.06.
\label{fig::compBCC}
}
\end{minipage}
\end{center}
\end{figure*}

\begin{figure*}[htb]
  \noindent
  \centering
\begin{tabular}{cccc}
  \multicolumn{4}{l}{a)} \\
    \multicolumn{4}{c}{\includegraphics*[width = 0.35 \textwidth]{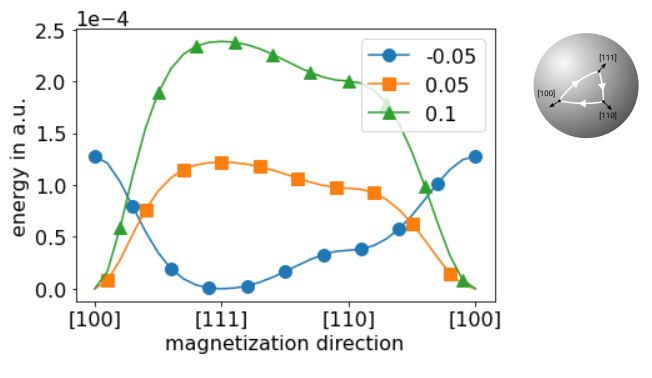}} \\
  \multicolumn{1}{l}{b)} & \multicolumn{1}{l}{c)} & \multicolumn{1}{l}{d)}&\\
  \includegraphics*[width=0.22\textwidth] {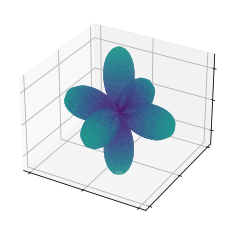} &
\includegraphics*[width=0.22\textwidth] {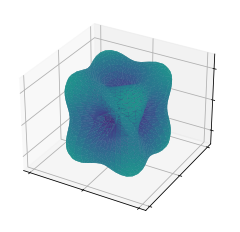} &
\includegraphics*[width=0.22\textwidth] {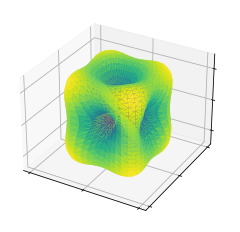}  & \includegraphics*[width=0.02\textwidth]{nice_plots/colorbar}\\
\aalpha=-0.05 & \aalpha=0.05 & \aalpha=0.1 &\\
\end{tabular}
\begin{center}
\begin{minipage}{0.9\textwidth}
\caption[short figure description]{
BCC, Model B. Comparing magnetic anisotropy for \aalpha = -0.05, 0.05 and 0.1. a) Energy along the edges of the triangle defined by \hkl[100], \hkl[111] and \hkl[110]. Maximum defines the hard and minimum the easy direction of magnetization. The energy is plotted relative to the minimum energy (easy direction). b)-d) Energy surface for \aalpha = -0.05, 0.05 and 0.1. The easy direction is \hkl[111] for \aalpha=-0.05 and \hkl[100] otherwise. 
\label{fig::compExtBCC}
}
\end{minipage}
\end{center}
\end{figure*}

The results reported in Fig.~\ref{fig:MESBCC} show the influence of \mm in \hkl[100] direction. The same analysis can be considered for other directions of \mm. Sampling all directions allows to compute the magnetic anisotropy. Figs. \ref{fig::compBCC} and \ref{fig::compExtBCC} show the energy for representative values of \aalpha for Model A and Model B, respectively.
For Model A the hard directions are always the \hkl<100> directions. The easy directions are \hkl<111>. Only for positive \aalpha the \hkl<110> becomes energetically comparable to \hkl<111>. However, we can conclude that for BCC crystals the hard direction of magnetization can not be controlled by Model A. This changes for Model B. With the considered set of parameters, for $\aalpha >0$ the easy directions of magnetization are \hkl<100> and the hard directions are \hkl<111>. 

\subsubsection{FCC crystals}

\begin{figure*}[htb]
  \noindent
  \centering
\begin{tabular}{cccc}
  \multicolumn{4}{l}{a)} \\
    \multicolumn{4}{c}{\includegraphics*[width = 0.35 \textwidth]{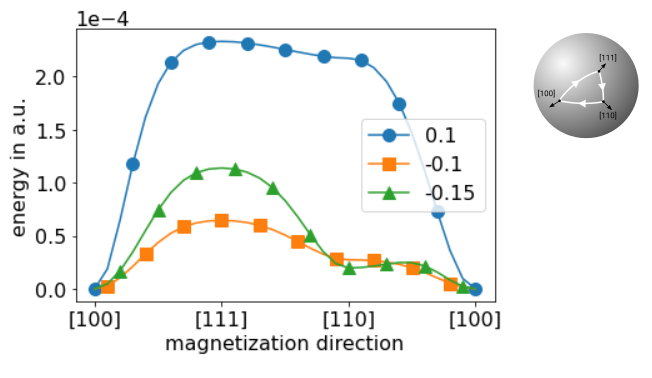}} \\
  \multicolumn{1}{l}{b)} & \multicolumn{1}{l}{c)} & \multicolumn{1}{l}{d)} &\\
\includegraphics*[width=0.22\textwidth] {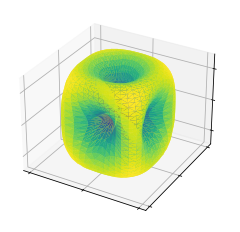} &
\includegraphics*[width=0.22\textwidth] {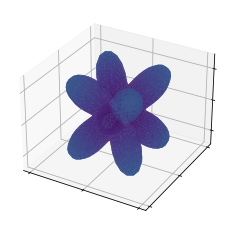} &
\includegraphics*[width=0.22\textwidth] {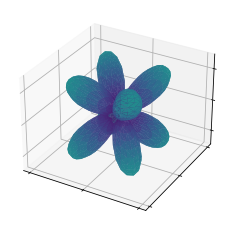} & \includegraphics*[width=0.02\textwidth]{nice_plots/colorbar} \\
\aalpha=0.1 & \aalpha=-0.1 & \aalpha=-0.15 & \\
\end{tabular}
\begin{center}
\begin{minipage}{0.9\textwidth}
\caption[short figure description]{
FCC, Model A. Comparing magnetic anisotropy for \aalpha = 0.1, -0.1 and -0.15. a) Energy along the edges of the triangle defined by \hkl[100], \hkl[111] and \hkl[110]. Maximum defines the hard direction and minimum the easy direction of magnetization. The energy is plotted relative to the minimum energy (easy direction). b)-d) Energy surface for \aalpha = 0.1, -0.1 and -0.15.  
\label{fig::compFCC}
}
\end{minipage}
\end{center}
\end{figure*}

\begin{figure*}[htb]
  \noindent
  \centering
\begin{tabular}{cccc}
  \multicolumn{4}{l}{a)} \\
    \multicolumn{4}{c}{\includegraphics*[width = 0.35 \textwidth]{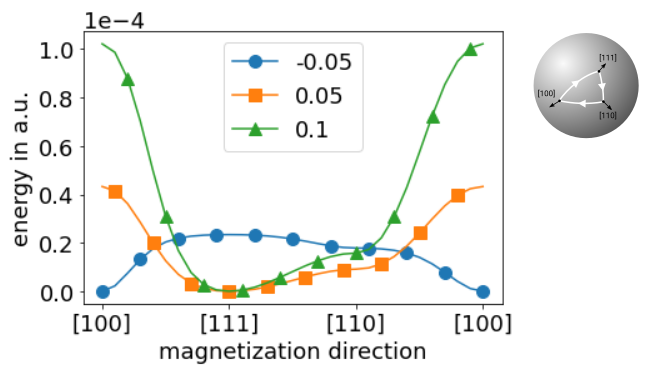}} \\
  \multicolumn{1}{l}{b)} & \multicolumn{1}{l}{c)} & \multicolumn{1}{l}{d)} & \\
  \includegraphics*[width=0.22\textwidth] {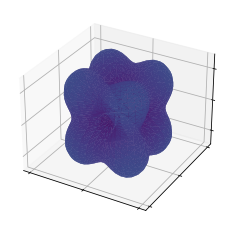} &
\includegraphics*[width=0.22\textwidth] {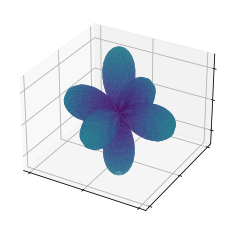} &
\includegraphics*[width=0.22\textwidth] {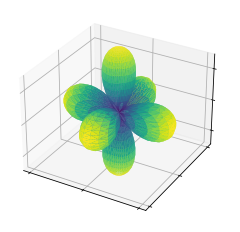}& \includegraphics*[width=0.02\textwidth]{nice_plots/colorbar}  \\
\aalpha=-0.05 & \aalpha=0.05 & \aalpha=0.1&  \\
\end{tabular}
\begin{center}
\begin{minipage}{0.9\textwidth}
\caption[short figure description]{
FCC, Model B. Comparing magnetic anisotropy for \aalpha = -0.05, 0.05 and 0.1. a) Energy along the edges of the triangle defined by \hkl[100], \hkl[111] and \hkl[110]. Maximum defines the hard direction and minimum the easy direction of magnetization. The energy is plotted relative to the minimum energy (easy direction). b)-d) Energy surface for \aalpha = -0.05, 0.05 and 0.1. The easy direction is \hkl[111] for \aalpha=-0.05 and \hkl[100] otherwise. 
\label{fig::compExtFCC}
}
\end{minipage}
\end{center}
\end{figure*}

For FCC crystals the structure dependent part in eq.~\eqref{eq:APFCgS} reads
\begin{align}
  f^{\mathrm{FCC}}(\sofA) =& -2t[\A_1^*(\A_2^*\A_5+\A_3^*\A_7+\A_4^*\A_6^*)+\A_2^*(\A_3^*\A_6+\A_4^*\A_7^*) +\A_3^*\A_4^*\A_5^*+{\rm {c.c.}}] \nonumber \\
& + 6v [\A_1^*(\A_2^*\A_3^*\A_4^*
+\A_2 \A_6^* \A_7  + \A_3\A_5 \A_6^* + \A_4\A_5\A_7) + \A_2^*\A_5(\A_3\A_7^*+\A_4\A_6) + \A_3^*\A_4\A_6\A_7+{\rm c.c.}].
                            \label{eq::gFCC}
\end{align}
and $\kkj$ vectors are defined as, see e.g. \cite{SBE17},
\begin{align}
  \sof[\kkj]&= \frac{1}{\sqrt{3}} \left[ \mycvec{-1\\ 1\\ 1},
  \mycvec{1\\-1\\1},\mycvec{1\\1\\-1},  \mycvec{-1\\-1\\-1}, \mycvec{2\\0\\0}, \mycvec{0\\2\\0},\mycvec{0\\0\\2} \right].
  \label{eq::kjsFCC}
\end{align}
They are numbered from $j = 1, \ldots, 7$ and shown in Fig.~\ref{fig::kjs}~b). Due to their different length they define two \MES. The amplitude expansion for this setting follows
from PFC models with two or more modes, e.g.~\cite{MEH13, YHT10, GPR10, WAK10}. However, in eq.~\eqref{eq::APFCenergy} we can account for this feature by considering different values for $q_0$ for the different sets of $\kkj$ vectors, see  \cite{EHP10,WAK10}. Eq.~\eqref{eq::FMAenergy} thus reads
\begin{equation}
\begin{split}
  \!\!\!\!\FF[\sofA,\{\kkj^\prime\}]= |\Omega^\prime| \Big\{g^{\mathrm{S}}(\sofA)+&\sum_{j=1}^4 \Aj B_0^x\left[\left(1-{\kkj^\prime}^2\right)^2 \!\!\!- 
  \alpha_{2} (\mm \cdot \kkj^\prime)^2 \right] \! \Aj \\ + &\sum_{j=5}^7 \Aj B_0^x\left[\left(\frac{4}{3}-{\kkj^\prime}^2\right)^2 \!\!\!- 
  \alpha_{2} (\mm \cdot \kkj^\prime)^2 \right] \! \Aj \Big\}, 
  \label{eq::FMAenergyFCC}
  \end{split}
\end{equation}
with $q_0 = \{1, \frac{4}{3}\}$ for the different sets of $\kkj$'s in Model A and analogously in Model B. Model B now allows to tune the parameter $q_m$ independently for the different sets of $\kkj$'s. We consider $q_m = \{1/\sqrt{3}, 1/\sqrt{2}\}$ for $\kk_{1-4}$ and $\kk_{5-7}$, respectively. With these modifications the same analysis as for BCC crystals can be done. Figure \ref{fig::compFCC} and \ref{fig::compExtFCC} show the results concerning mangetic anisotropy for Model A and Model B, respectively.

For FCC, Model A describes the easy directions always aligned tothe \hkl<100> directions. The hard directions are \hkl<111>. Only for highly positive \aalpha the \hkl<110> becomes energetically comparable to \hkl<111>. However, we can conclude that for FCC crystals the easy direction of magnetization cannot be controlled by Model A. As for BCC this changes for Model B. With the considered set of parameters for $\aalpha >0$ the easy directions of magnetization are \hkl<111> and the hard directions are \hkl<100>.

\subsubsection{Magnetic anisotropy of ferromagnetic materials}
Tuning the easy and hard directions of magnetization becomes necessary as the magnetic anisotropy of various ferromagnetic materials features \hkl[100] and \hkl[111] as the easy direction of magnetization for BCC and FCC crystals, respectively. Model A does not allow for such versatility. Thus, to account for the proper easy and hard directions of magnetization, Model B has to be considered. Table \ref{tab::matAniso} shows such directions for Fe, Ni, and Co. 

\begin{table}[htb]
  \begin{tabular}{ccccc}
    material & crystal & easy & hard & Ref.\\
    & structure &  \multicolumn{2}{c}{directions} & \\ \hline
    Fe & BCC & \hkl<100>& \hkl<111> & \cite{HPA98} \\
    Ni & FCC & \hkl<111>& \hkl<100> & \cite{YNM94,HPA98} \\
    Co & FCC & \hkl<111>& \hkl<100> & \cite{HPA98}
  \end{tabular}
  \begin{minipage}{0.9\textwidth}
\caption[short figure description]{Material specific magnetic anisotropies. 
\label{tab::matAniso}
}
\end{minipage}
\end{table}  

These properties also hold for various alloys, e.g. Fe$_{1-x}$Ga$_x$ \cite{RCC04}. PFC and APFC models for alloys have been introduced in \cite{Huang10}, and magnetic coupling for these models has been introduced in \cite{SSP15}. The considered modifications in Model B can also be applied to these models.



\subsection{Impact of magnetization on grain growth}
\begin{figure*}[htb]
  \noindent
  \centering
  \begin{tabular}{ccc}
    \multicolumn{1}{l}{a)} & \multicolumn{1}{l}{b)} & \multicolumn{1}{l}{c)} \\ 
    \raisebox{.0\height}{\includegraphics*[ width = 0.2\textwidth ]{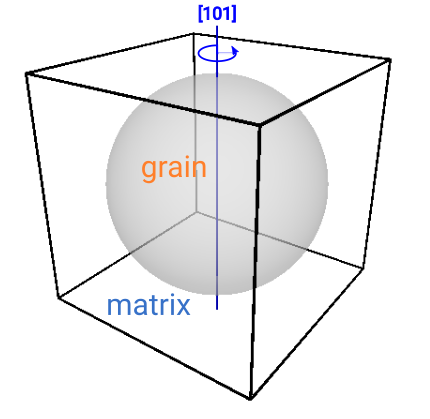}} &
    \raisebox{.0\height}{\includegraphics*[ width = 0.2\textwidth ]{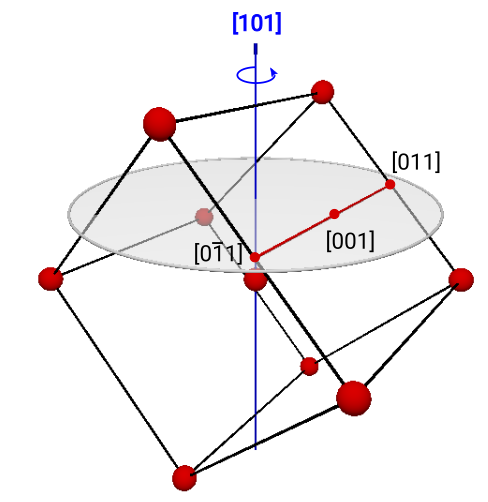} } & \hspace*{0.4cm}
    \raisebox{.0\height}{\includegraphics*[ width = 0.25\textwidth ]{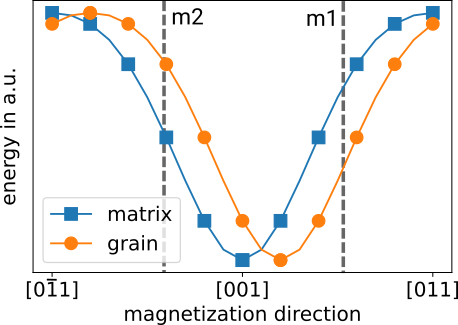}} 
  \end{tabular}
  \begin{center}
\begin{minipage}{0.9\textwidth}
\caption{
Setup for magnetic APFC simulation of grain shrinkage. a) Schematics of the spherical grain rotated about the \hkl[101] axis (grain) and embedded in an unrotated crystal (matrix). b) Illustration of the crystals unit cell: the \hkl[101] direction of both the grain and matrix is aligned with the axis of rotation. The direction of magnetization is defined w.r.t the crystallographic orientations of the matrix. c) Magnetic anisotropy of matrix (blue squares) and grain (orange circles) along the lines defined by \hkl[0-11], \hkl[001] and \hkl[011]. 
m1 and m2 point at specific magnetizations considered in simulations, favoring the crystal structure in the grain and the matrix, respectively, while maximising the corresponding energy difference.
\label{fig::APFCsetUp}
}
\end{minipage}
\end{center}
\end{figure*}

We consider a system with the basic magnetic properties of Fe and examine the influence of magnetization on the shrinkage of an initially spherical grain with a small rotation with respect to the surrounding matrix \cite{YMV17,SBV18,SVE19}. 
We consider a BCC crystal and Model B with \aalpha = 0.1 and $q_m = 1/\sqrt{2}$. The spherical grain has a radius of $60 \pi$, and is rotated about the \hkl[101] direction by $5^\circ$ with respect to the surrounding matrix, see Fig.~\ref{fig::APFCsetUp}~a). 
This initial setting is considered by a definition of the phase of complex amplitudes in the grain reading $\delta \mathbf{k}(\theta)\cdot \mathbf{r}$ and vanishing in the matrix, with $\delta \mathbf{k}(\theta)$ the difference between the rotated and unrotated \kkj vectors (see \cite{SBE17,SBV18,SVE19} for more details).
Fig.~\ref{fig::APFCsetUp}~b) illustrates the cubic unit cell. The rotational axis, \hkl[101], coincides with the grain and the matrix, while magnetizations are defined with respect to the crystallographic axes of the matrix. For a magnetization oriented along the \hkl[0-11], \hkl[001], and \hkl[011] directions, the free energy of the matrix and grain, computed as reported in the previous sections, vary as illustrated in Fig.~\ref{fig::APFCsetUp}~c). For the matrix, there are energy maxima for \mm oriented along \hkl<011> directions and energy minima for \mm oriented along \hkl<001> directions. For the rotated grain, magnetization in these directions results in slightly shifted easy and hard directions, owing to the (small) rotation of its crystallographic axes. Also, the energy difference between grain and matrix varies with the orientation of \mm. We select orientations of \mm, which maximize this difference. In particular, we consider \mm between \hkl[001] and \hkl[011], (m1), for which the crystal structure in the grain is energetically favorite, and between \hkl[0-11] and \hkl[001], (m2), for which the crystal structure in the matrix is favorite, see Fig.~\ref{fig::APFCsetUp}~c). No preference for grain or matrix is achieved by choosing \mm along the axis of rotation of the grain, \hkl[101], m0, which will be considered for comparison.

\begin{figure*}[htb]
  \noindent
  \centering
    \includegraphics*[ width = 0.99\textwidth ]{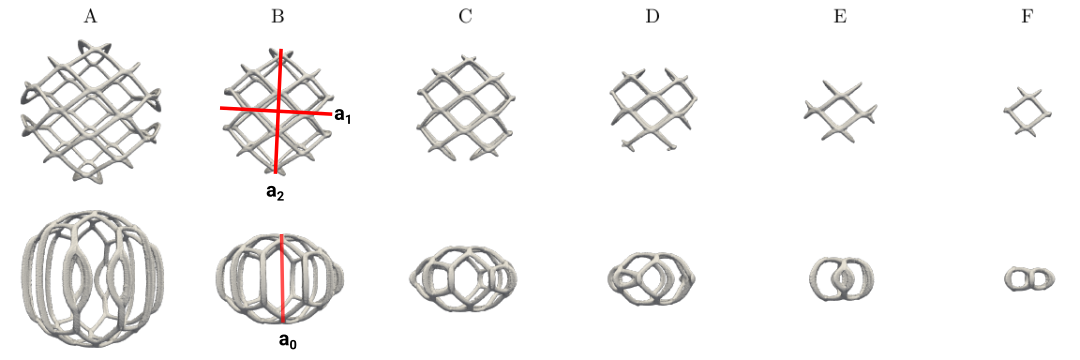} 
  \begin{center}
\begin{minipage}{0.9\textwidth}
\caption[short figure description]{
Defect networks of an initially spherical grain in a BCC crystal during shrinkage. A-F) are the structures are representative times, also indicated in Fig.~\ref{fig::evalShrinkage}. The magnetization is aligned with the axis of rotation (m0). Measures of the grain extension, $\rm a_{0-2}$, are introduced to characterize its shape, corresponding to axes of an ellipsoid approximating the grain-matrix interface extension. $\rm a_0$ is parallel to the rotation axis or \hkl[101], see Fig.~\ref{fig::APFCsetUp}. $\rm a_{1,2}$ lay in the corresponding planes.
\label{fig::shrinkage}
}
\end{minipage}
\end{center}
\end{figure*}

Fig.~\ref{fig::shrinkage} illustrates the grain shrinkage for the magnetization m0. In particular, the dislocation network forming between grain and matrix is shown. This is obtained by exploiting the decrease of amplitudes at defects. We consider here regions with $A^2<0.083$ see \cite{SBV18,Salvalaglio2022}. The dislocation network shrinks anisotropically, as it has been observed in previous studies \cite{YMV17,SBV18}. For m0, m1, and m2, shrinking grains with very similar dislocation networks (as in Fig.~\ref{fig::shrinkage}) are obtained, indicating negligible effects on the fine details of the dislocation network structure. This can be ascribed to the unchanged incommensurability of the crystals in the grain and the matrix, leading to similar topological defects and small changes in the elastic interactions due to small magnetostriction as dictated by model parameters and magnetization direction. Qualitatively different results are indeed obtained if considering Model A, resulting however from an unphysically large magnetostriction and with easy and hard directions that cannot be tuned \cite{Backofen2022}. Importantly, the shrinkage speed is largely influenced by the orientation of \mm, which due to magnetic anisotropy introduces an additional driving force related to the differences in the bulk energy when moving across the grain boundary between grain and matrix.

\begin{figure}[htb]
\noindent
\begin{tabular}{ccc}
   \multicolumn{2}{l}{a)} \\
   \multicolumn{2}{c}{\includegraphics*[width = 0.42 \textwidth]{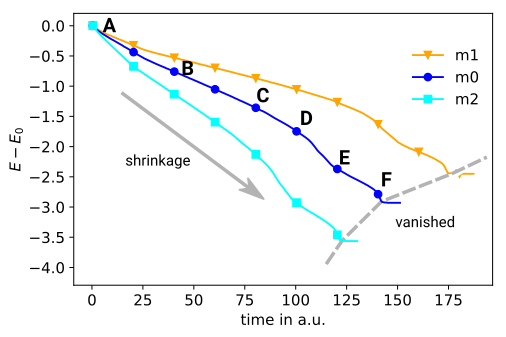}} \\
       \multicolumn{1}{l}{b)} &\multicolumn{1}{l}{c)} \\
    \includegraphics*[ width = 0.42\textwidth ]{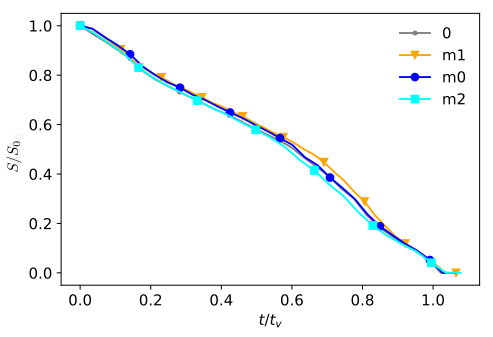} &
      \includegraphics*[ width = 0.42 \textwidth ]{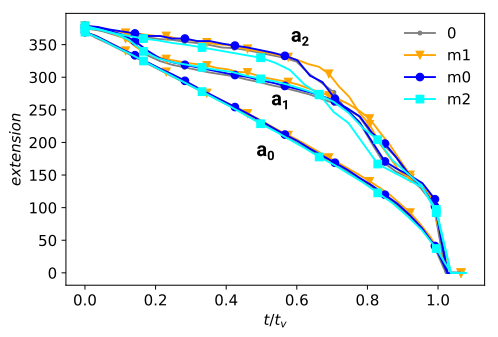} \\
  \end{tabular}
\begin{center}
\begin{minipage}{0.95\textwidth}
\caption[short figure description]{
  Influence of magnetization on grain shrinkage. a) Energy decay during shrinkage. Dependent on magnetization the grain vanishes at different times $t_v$. b) Grain shrinkage illustrated by the decrease over time of the grain boundary between grain and matrix, which is approximated by Eq.~\eqref{eq:KT} normalized by the initial area. The timescale is normalized by the vanishing time $t_v$. c)  Decrease over time of $\rm a_{0,1,2}$ (as defined in Fig.~\ref{fig::shrinkage}).  
\label{fig::evalShrinkage}
 }
\end{minipage}
\end{center}
\end{figure}

A detailed analysis of this evidence is reported in Fig.~\ref{fig::evalShrinkage}. Fig.~\ref{fig::evalShrinkage}~a) shows the energy decay relative to the initial energy during grain shrinkage. Constant energy is obtained when the grain vanishes, at a time here referred to as vanishing time $t_v$, which is found to depend on the magnetization. The slowest shrinkage is achieved if the grain is energetically preferred, m1. Here the additional driving force related to magnetization tends to favor the crystal structure in the grain, opposing the shrinkage dynamics (well described by mean-curvature flow \cite{DHR97}). The energy gain during shrinkage is the lowest. Consistently, the highest shrinkage speed and energy gain are observed when the grain has higher energy, m2. Here the driving force introduced by magnetization enhances the underlying shrinkage dynamic favoring the crystal structure in the matrix. If matrix and grain are energetically equivalent, m0, the vanishing time is between the two cases. Thus, the magnetization enhances or hinders grain shrinkage according to its direction. 

More details on the structure of the evolving dislocation network are reported in Fig.~\ref{fig::evalShrinkage}~b) and c).
The initially spherical small-angle (semi-coherent) grain boundary is approximated as the surface of a three-dimensional ellipsoid interpolating the dislocation network, whose axes are along $x$,$y$, and $z$-direction are denoted as $\rm a_{0,1,2}$. The corresponding surface area is computed by the so-called Knud Thomsen's formula \cite{SBV18}:
\begin{eqnarray}\label{eq:KT}
  S=\frac{\pi}{3^{1/p}} \left( ({\rm a_0 a_1})^p+({\rm a_1 a_2})^p+({\rm a_0 a_2})^p \right)^{1/p},
\end{eqnarray}
with $p=1.6075$. This quantity normalized by the surface area of the initial spherical grain as well as the axes $\rm a_{0,1,2}$ are reported in Fig.~\ref{fig::evalShrinkage}~b) and c), respectively, against the time normalized by $t_v$. In all considered cases, the grain boundary between grain and matrix decreases nearly linearly (Fig.~\ref{fig::evalShrinkage}~b). Such a linear scaling, as well as a linear decrease in energy, is predicted by the classical theory of grain shrinkage driven by mean curvature flow \cite{DHR97}, and it is reproduced by PFC and APFC without magnetic interaction \cite{YMV17,SBV18}. This indicates that the evolution is still mainly governed by the minimization of interface energy, owing to the relatively small size of the grain and, thus, the relatively large mean curvature of the grain boundary between grain and matrix. Deviations are expected for larger systems \cite{SSH2022}. Nevertheless, the magnetic anisotropy may enhance or hinder the evolution with a volumetric-energy contribution that acts isotropically on the grain boundary and its dislocation network. This is further shown in Fig.~\ref{fig::evalShrinkage}~c) where the evolution of the axes $\rm a_{0,1,2}$ are shown. $\rm a_0$ shrinks nearly linearly with a constant speed up to the end of the shrinking process, $t/t_v>0.9$. A more complex evolution is observed for $\rm a_1$ and $\rm a_2$ due to defect annihilation. In particular, in the interval $0.6<t/t_v<0.8$ indicated by C-E in Fig.~\ref{fig::shrinkage}, dislocation lines vanish, and sudden changes in the dislocation network morphology occur. This stage can also be detected in the decay of energy and grain boundary surface, where the decay rate deviates the most from linear scaling and could be ascribed to additional elasticity effects enhanced by short-range dislocation interaction. However, the magnetization is found to affect negligibly the anisotropy during the shrinkage of the grain as $\rm a_{0,1,2}$ vary only slightly among the different chosen orientations of $\mm$. 

\section{Conclusion} \label{sec:4}
We have reviewed and extended magnetic PFC and APFC models. Our focus has been on the control of magnetic anisotropy in these models. For various ferromagnetic materials, the easy direction of magnetization is \hkl[100] for BCC crystals and \hkl[111] for FCC crystals. Modeling this behavior requires an extension of magnetic couplings in existing models. By analyzing the Minimal Energy Surface (\MES), we explored the possibility of tuning the easy and hard direction of magnetization by including higher-order coupling terms. This can be achieved without increasing the complexity of the model significantly. The numerical realization only requires directional derivatives of order four and thus does not increases the order of the derivatives in the equations. The higher-order terms describe a double well in the direction of magnetization in the reciprocal space. Therefore, the local extrema
of the double well can be chosen freely, and the energy contributions at the
\MES are changed without deformation of its shape. The considered parameters are the coupling strength \aalpha and an additional length scale $q_m$. Both can be used to tune the magnetic anisotropy to those of specific ferromagnetic materials. 

Besides the influence of magnetic anisotropy, the magnetic coupling terms also influence magnetostriction. Both phenomena depend on the coupling strength \aalpha and are strongly correlated. However, in some cases, the magnetostriction vanishes, and Model B allows for controlling this through $q_m$. A different approach to decoupling both phenomena is addressed in \cite{BV22}.

The model has been applied to the simulation of the shrinkage of a spherical grain in a matrix under the influence of a constant magnetic field and using the basic magnetic properties of Fe. 
The shrinkage is anisotropic and can be enhanced or hindered by magnetization. However, the details of the considered magnetic coupling affects the shrinkage only slightly in terms of morphologies of dislocation networks and scaling laws. This is attributed to the small grain and, thus, dominating effects of the curvature of the grain boundary between grain and matrix.

\section*{Acknowledgements}
AV and RB acknowledge support by the German Research Foundation (DFG) within SPP1959
under Grant No. VO899/20-2. MS acknowledges support from the Emmy Noether Programme of the German Research Foundation (DFG) under Grant No. SA4032/2-1. We further
acknowledge computing resources provided at the Center for Information Services and HPC (ZIH) - TU Dresden, and J\"ulich Supercomputing Center under Grant
PFAMDIS.

\bibliography{allDiss}

\begin{figure}
\textbf{Table of Contents}\\
\medskip
  \includegraphics{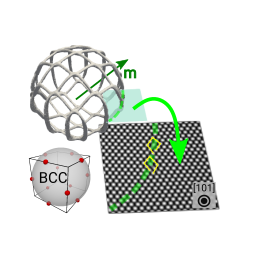}
  \medskip
  \caption*{How do magnetic fields interact with dislocations and what is the effect of this interaction on the microstructure of ferromagnetic materials?
A multiscale modeling approach is considered which allows to answer these questions. Parameterised for Fe the influence of a magnetic field
on the evolution of the dislocation network of a spherical grain is analysed. 

 }
\end{figure}

\end{document}